\documentclass[journal,twoside,web]{ieeecolor}
\usepackage{generic}
\usepackage{cite}
\usepackage{amsmath,amssymb,amsfonts}
\usepackage{algorithmic}
\usepackage{graphicx}
\usepackage{algorithm,algorithmic}
\usepackage{hyperref}
\hypersetup{hidelinks=true}
\usepackage{textcomp}

\usepackage{booktabs,multirow}

\def\BibTeX{{\rm B\kern-.05em{\sc i\kern-.025em b}\kern-.08em
    T\kern-.1667em\lower.7ex\hbox{E}\kern-.125emX}}
\begin{document}
\title{Multilevel Perception Boundary-guided Network for Breast Lesion Segmentation in Ultrasound Images}
\author{Xing Yang, Jian Zhang, Qijian Chen, Li Wang and Lihui Wang, \IEEEmembership{Member, IEEE}
\thanks{Manuscript received xxxx; accepted xxxx. Date of publication November xxxx; date of current version xxxx. This work was supported by the National Natural Science Foundation of China (Grant No.62161004), Guizhou Provincial Science and Technology Projects (QianKeHe ZK [2021] Key 002), the Nature Science Foundation of Guizhou Province (QianKeHe [2020]1Y255), and Guizhou Provincial Science and Technology Projects (QianKeHe ZK [2022] 046). (Corresponding author: Jian Zhang) }
\thanks{Xing Yang, Jian Zhang, Qijian Chen, Li Wang and Lihui Wang are with the Engineering Research Center of Text Computing \& Cognitive Intelligence, Ministry of Education, Key Laboratory of Intelligent Medical Image Analysis and Precise Diagnosis of Guizhou Province, State Key Laboratory of Public Big Data, College of Computer Science and Technology, Guizhou University, Guiyang 550025, China (e-mail: gs.xingyang21@gzu.edu.cn, jzhang7@gzu.edu.cn, forchenqijian@163.com, wangli614079@163.com, lhwang2@gzu.edu.cn).}
\thanks{Digital Object Identifier xxxx}}
\maketitle

\begin{abstract}
Automatic segmentation of breast tumors from the ultrasound images is essential for the subsequent clinical diagnosis and treatment plan. Although the existing deep learning-based methods have achieved significant progress in automatic segmentation of breast tumor, their performance on tumors with similar intensity to the normal tissues is still not pleasant, especially for the tumor boundaries. To address this issue, we propose a PBNet composed by a multilevel global perception module (MGPM) and a boundary guided module (BGM) to segment breast tumors from ultrasound images. Specifically, in MGPM, the long-range spatial dependence between the voxels in a single level feature maps are modeled, and then the multilevel semantic information is fused to promote the recognition ability of the model for non-enhanced tumors. In BGM, the tumor boundaries are extracted from the high-level semantic maps using the dilation and erosion effects of max pooling, such boundaries are then used to guide the fusion of low and high-level features. Moreover, to improve the segmentation performance for tumor boundaries, a multi-level boundary-enhanced segmentation (BS) loss is proposed. The extensive comparison experiments on both publicly available dataset and in-house dataset demonstrate that the proposed PBNet outperforms the state-of-the-art methods in terms of both qualitative visualization results and quantitative evaluation metrics, with the Dice score, Jaccard coefficient, Specificity and HD95 improved by 0.70\%, 1.1\%, 0.1\% and 2.5\% respectively. In addition, the ablation experiments validate that the proposed MGPM is indeed beneficial for distinguishing the non-enhanced tumors and the BGM as well as the BS loss are also helpful for refining the segmentation contours of the tumor. 
\end{abstract}


\section{Introduction}
\label{sec:introduction}
\IEEEPARstart{A}{ccording} to the World Health Organization (WHO), breast cancer has surpassed lung cancer as the most common cancer worldwide\cite{sung2021global}. 
Early diagnosis and precise treatment are critical to improving the survival rate of breast cancer patients\cite{chengApproachesAutomatedDetection2006}. 
Ultrasound imaging technology is widely used in early diagnosis of breast cancer due to its non-invasive, radiation-free, and low-cost characteristics\cite{xianAutomaticBreastUltrasound2018}. 
Accurate lesion segmentation in B-mode breast ultrasound (BUS) is essential for the subsequent diagnosis\cite{xianAutomaticBreastUltrasound2018}, treatment planning\cite{acostaEvaluationMultiatlasbasedSegmentation2011}, and prognostic assessment\cite{chenRandomWalkbasedAutomated2011}. 
Traditional BUS lesion segmentation is usually based on the subjective judgment and is highly dependent on the the experience of radiologists, which is time-consuming and labor-intensive due to the complexity of breast tumor shape and size. Therefore, investigating the accurate, automatic, and real-time breast ultrasound image tumor segmentation algorithms is of great significance.

Currently, automatic BUS segmentation methods can be divided into two categories: traditional ones and deep learning-based ones\cite{xianAutomaticBreastUltrasound2018}. 
The former includes threshold segmentation\cite{minFullyAutomaticSegmentation2015}, region growing\cite{kwakRDBasedSeededRegion2005}, watershed algorithm\cite{loMultiDimensionalTumorDetection2014}, deformable model\cite{weiFastSnakeModel2004}, and superpixel segmentation\cite{ilesanmiMultiscaleSuperpixelMethod2020} etc.
Even though these methods are easier to implement,  they require manual intervention for image processing, which lacks the adaptability and limits their performance in complex scenarios. In recent years, with the successful applications of deep learning in the filed of computation vision, several semantic segmentation architectures based on convolutional neural networks (CNN) have been proposed for BUS segmentation. For instance, Yap et al. \cite{yapAutomatedBreastUltrasound2018} first apply transfer learning techniques to localize breast cancer lesion areas using FCN-AlexNet\cite{longFullyConvolutionalNetworks2015a}. 
Vakanski et al. \cite{vakanskiAttentionEnrichedDeepLearning2020} use data augmentation strategies to train models, demonstrating the effectiveness of UNet\cite{olafUNetConvolutionalNetworks2015} and SegNet\cite{badrinarayananSegNetDeepConvolutional2017} in BUS segmentation. However, due to the large semantic gap between encoder and decoder features, feature fusion through skip connections makes it difficult to effectively transmit deeper semantic information. To address this problem, Ibtehaz et al.\cite{ibtehazMultiResUNetRethinkingUNet2020} embed several residual blocks at different levels of skip connections in UNet. To further improve the ability of network in capturing long-distance information, Byra et al. \cite{byraBreastMassSegmentation2020} use selective kernels\cite{liSelectiveKernelNetworks2019a} instead of regular convolution kernels to explore multi-scale spatial information more effectively and therefore to better segment breast cancer tumor areas of different sizes, it achieves  
an average Dice score of 79.1\% on the UDIAT dataset\cite{yapAutomatedBreastUltrasound2018}. However, this method only considers the interdependence of local spatial regions, which cannot capture contextual information in a global view. Accordingly, Vivek et al.\cite{singhBreastTumorSegmentation2020} use atrous convolution\cite{chenRethinkingAtrousConvolution2017} to capture image contextual information at multiple scales without reducing the sampling rate for segmentation, the average Dice score is increased to 86.82\% on the UDIAT dataset. Subsequently, Runze et al.\cite{wangMultilevelNestedPyramid2019} introduce atrous spatial pyramid pooling (ASPP) module \cite{chenDeepLabSemanticImage2018} into the skip connections to encode low-level features and high-level semantics in a multi-level and multi-scale manner for further improving the segmentation performance. In addition, considering that feature pyramid network (FPN) \cite{tsung-yiFeaturePyramidNetworks2017} is good at detecting targets of different sizes, Peng et al.\cite{tangFeaturePyramidNonlocal2021} combine FPN and self-attention mechanisms to segment the tumors with different shapes and sizes, which achieves a promising result.

Using the atrous convolution or pyramid pooling strategies can effectively deal with the segmentation task for breast tumors with different sizes, however, due to a large number of continuous pooling and convolution operations, image details may be lost, making it difficult for the model to accurately segment irregular tumor boundaries. To refine the segmented boundary, a lot of attention-based methods have been proposed. For instance, Fan et al. \cite{fanPraNetParallelReverse2020} design a parallel partial decoder (PraNet) that aggregates high-level features to generate a rough segmentation guide map, and then combine the guide map with the reverse attention (RA) \cite{chenReverseAttentionSalient2018} module to establish the relationship between the segmentation region and the boundaries, from such relationship, the detailed lesion boundaries can be extracted from the low-level features of lesion region. Inspired by this work,  Lou et al. \cite{louCaraNetContextAxial2022} propose an axial reverse attention (A-RA) module which combines the axial and reverse attentions to obtain more precise feature information for improving the segmentation performance on boundaries. Hu et al. \cite{huBoundaryguidedRegionawareNetwork2023} propose a boundary-guided and region-aware network to eliminate the influence of irrelevant backgrounds(i.e., focus only on tumor areas) and to detect tumor boundaries. This work can improve the segmentation performance on the tumor boundaries, but it increases the computational complexity, in addition, it does not fully consider the relationship between the tumor, normal, and boundary regions, which may influence the segmentation performance.

Although the above-mentioned methods can deal with respectively the breast tumors with different sizes and irregular shapes, due to the presence of many low echogenic areas\cite{minFullyAutomaticSegmentation2015} in BUS that are extremely similar to lesion regions, and the high variability of tumors in shape, size, appearance, texture, and location, segmenting accurately the heterogenous breast tumors from BUS images is still challenging. To address this challenge, we propose a novel multilevel perception boundary-guided network (PBNet) for breast tumor segmentation, which consists of a multilevel global perception module (MGPM) and a boundary guided module (BGM). MGPM is responsible for extracting the segmentation-related features by integrating the long-range dependency information at the same semantic level as well as at higher semantic levels. BGM is used to assist in extracting the lesion boundary information by generating the boundary-guided attention map. In addition, to further enhance the ability of PBNet in localizing lesion boundaries, we design also a boundary-enhanced segmentation (BS) loss. With the help of MGPM, BGM and BS loss, our PBNet can achieve better performance than state-of-the-art (SOTA) methods.

The rest of this paper is organized as follows: section \ref{sec:methods} demonstrates the details of the proposed PBNet, section \ref{sec:experiments} describes the experimental setups, experimental results and discussion are demonstrated in sections \ref{sec:results} and \ref{sec:discussion}, respectively, and section \ref{sec:conclusion} makes a conclusion about this work.

\section{Methods}
\label{sec:methods}

\begin{figure}[!t]
	\centering
	\includegraphics[width=1\columnwidth]{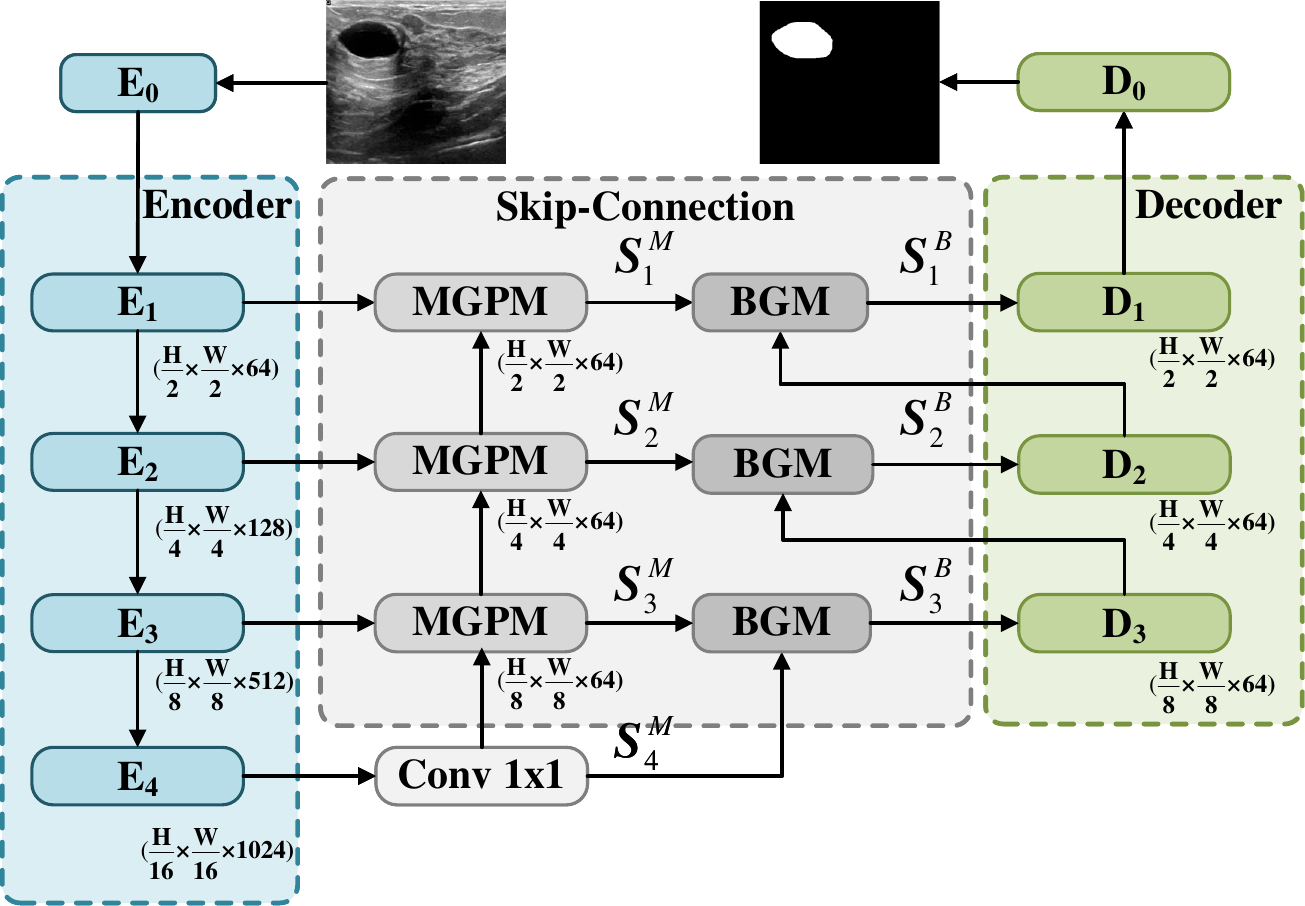}
	\caption{Overview structure of PBNet. MGPM indicates multilevel global perception module and BGM represents boundary-guided module.}
	\label{fig:pbnet}
\end{figure}

PBNet has a typical encoder-decoder architecture, as shown in Fig. \ref{fig:pbnet}. We use EfficientNet \cite{tanEfficientNetRethinkingModel2020} to extract multi-level feature maps $E_l, (l=0,1,...4)$ with different spatial resolutions. To reduce model complexity, we use a $1 \times 1$ convolution in the skip connection to change the number of channels of each encoder feature $E_l$ to $64$. To fully utilize the complementary information between semantic features at different levels, as well as to enhance the model's ability to learn long-range information, and to improve its awareness of tumor boundaries, we introduce MGPM and BGM. The decoder infers the final segmentation result from the output feature maps of BGM. In the following sections, we will introduce the specific structure of MGPM and BGM in detail.

\subsection{Multilevel Global Perception Module}
The MGPM aims to enhance the ability of the model in capturing contextual information, and accordingly to improve the segmentation accuracy of non-enhanced tumors. To achieve this, MGPM integrates low-level semantic features $E_{l}$ and high-level ones $S_{l+1}^M$ through a multi-scale communication mechanism.

\begin{figure}[!t]
	\centering
	\includegraphics[width=1\columnwidth]{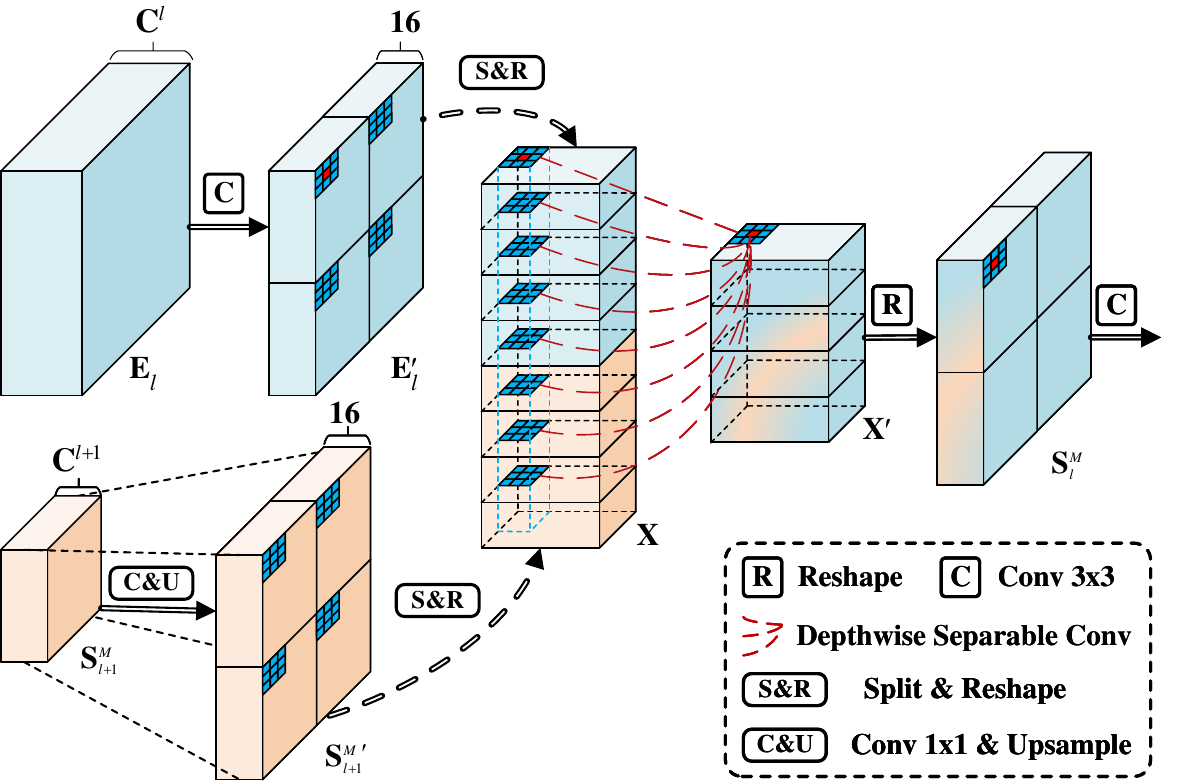}
	\caption{The schematic illustration of the multilevel global perception module (MGPM), where ${E_l}$ is the output feature map of Encoder at layer $l$ and ${S_{l+1}^M}$ is the output feature map of MGPM at layer ${l+1}$.}
	\label{fig:mgpm}
\end{figure}

As shown in Fig. \ref{fig:mgpm}, for the input low-level semantic features ${E_l}\in {{\mathbb{R}}^{{{C}^{l}}\times {{H}^{l}}\times {{W}^{l}}}}$, we use a $1\times 1$ convolution to reduce its number of channels to $32$, noting as ${{E}_{l}^{\prime}}\in {{\mathbb{R}}^{{32}\times {{H}^{l}}\times {{W}^{l}}}}$. For the high-level features ${S_{l+1}^M}\in{{\mathbb{R}}^{C^{l+1}\times H^{l+1}\times W^{l+1}}}$, they are first upsampled with a factor of 2, and then their channels are also reduced to $32$ with a $1 \times 1$ convolution, the resulted feature maps are noted as ${S_{l+1}^{M\prime }}\in{{\mathbb{R}}^{32\times H^{l}\times W^{l}}}$. 
Then, inspired by AFNet\cite{fengAttentiveFeedbackNetwork2019}, we split the low-level feature$E_l^{\prime}$ and the high-level features $S_{l+1}^{M\prime}$ into  $n\times n$  cells (in this paper, $n=2$), and each cell has a semantic feature map size of $({{H}^{l}}/n,{{W}^{l}}/n)$. The split high-level and low-level semantic feature maps are concatenated along the channel dimension to obtain a feature map ${X}\in {{\mathbb{R}}^{(2\times {{n}^{2}}\times 16)\times ({{H}^{l}}/n)\times ({{W}^{l}}/n)}}$. 
Subsequently, a fused feature $X^{\prime}\in {{\mathbb{R}}^{({{n}^{2}}\times 16)\times ({{H}^{l}}/n)\times ({{W}^{l}}/n)}}$ is obtained through a $3\times 3$ depth-wise separable convolution. After the fusion, each target pixel (shown in red) in $X^{\prime}$ is related to the other ${{3}^{2}}\times {2\times({n}^{2}}-1)$ pixels (shown in blue). This integrates neighboring and distant pixel information at the same level as well as the distant pixel information at higher levels. Therefore, MGPM module can expand the receptive field and increase the ability to capture contextual information. 
Finally, the feature map $X^{\prime}$ is reshaped to size $(32\times H^l\times W^l)$. After two $3\times3$ convolutions, the number of channels is increased to $64$ to obtain the final output of MGPM $S_l^M\in {{\mathbb{R}}^{64\times H^l \times W^l}}$.

As shown in Fig. \ref{fig:pbnet}, when $l=3$, the high-level semantic feature ${S_{l+1}^M}$ is obtained from the low-level semantic feature ${E_{l+1}}$ through a $1 \times 1$ convolution. Note that there is a BatchNorm layer and a ReLU activation function after each convolution.

\subsection{Boundary-guided Module}
\begin{figure*}[t]
	\centering
	\includegraphics[width=0.7\textwidth]{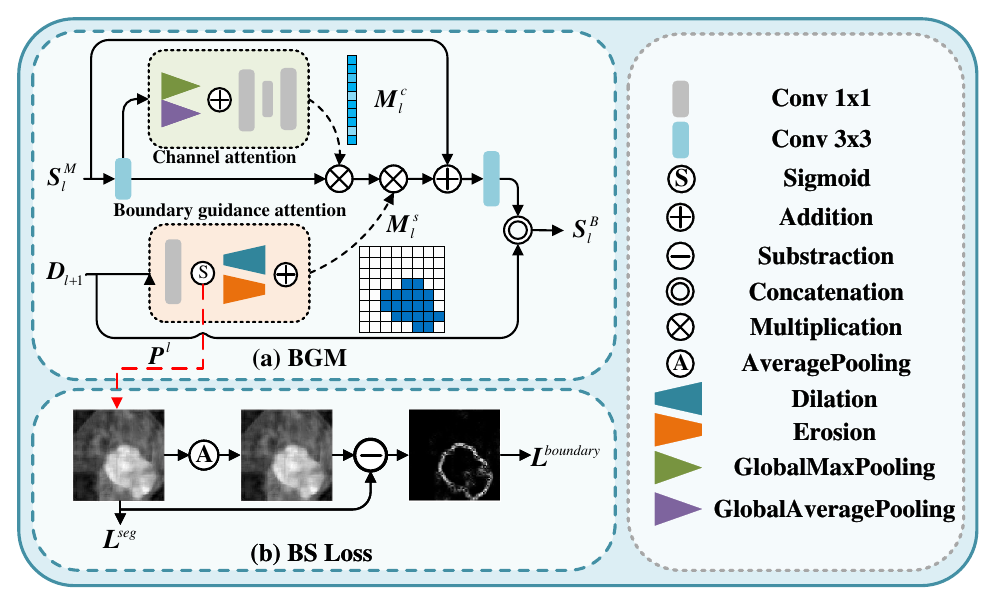}
	\caption{The architecture of the boundary-guided module (BGM), where ${S_{l}^M}$  is the output feature map of MGPM at layer ${l}$ and ${D_{l+1}}$ is the output feature map of decoder at layer ${l+1}$.}
	\label{fig:bgm}
\end{figure*}

Although MGPM is able to capture long-distance contextual information that is beneficial for distinguishing low-echogenic areas and tumor regions in breast ultrasound images, the model still cannot accurately localize tumor boundaries. To address this issue, we propose the BGM to further refine the output feature map of MGPM and improve the segmentation accuracy of tumor boundaries. As shown in Fig. \ref{fig:bgm}, the input of BGM at the $l^{th}$ level includes the output $S_l^M$ of the $l^{th}$ MGPM and the output of decoder $D_{l+1}$ at $(l+1)^{th}$ level. The corresponding output feature map $S_l^B$ of BGM can be expressed as:
\begin{equation}
	\label{eq:slb}
	\begin{aligned}
		{S_l^B}= & \mathop{\mathbf{Conv}}\limits_ {3\times3}(M_l^s \times(M_l^c\times \mathop{\mathbf{Conv}}\limits_ {3\times3}(S_l^M)) \oplus S_l^M)\\
		& || (\uparrow D_{l+1}), \quad l=1,2,3.
	\end{aligned}	
\end{equation}
where $\mathop{\mathbf{Conv}}\limits_ {3\times3}$ represents a $3\times 3$ convolution layer, $M_l^c$ represents channel attention map, $M_l^s$ represents boundary-guided spatial attention map, $\oplus$ and $||$ represent pixel-by-pixel addition and concatenation operations, respectively, and $\uparrow$ represents $\times 2$ upsampling. When $l=3$, $D_{l+1}=\mathop{\mathbf{Conv}}\limits_{1\times 1}(E_{l+1})$.

As shown in Fig. \ref{fig:bgm}, $M_l^c$ is derived from the channel attention block, in which, the global average-pooling and global max-pooling are performed respectively on the output feature maps 
${S_l^M}$ of MGPM to generate two different spatial context feature vectors ${{F}'}_{avg}^{{}}\in {{\mathbb{R}}^{C\times 1\times 1}}$ and ${{F}'}_{max}^{{}}\in {{\mathbb{R}}^{C\times 1\times 1}}$. They are then fed into a parameter-shared MLP to obtain new feature vectors ${F}_{avg}\in {{\mathbb{R}}^{C\times 1\times 1}}$ and ${F}_{max}\in {{\mathbb{R}}^{C\times 1\times 1}}$. Finally, ${{F}_{avg}}$ and ${{F}_{max}}$ are summed and passed through a sigmoid activation function to obtain the channel attention map $M_l^c\in {{\mathbb{R}}^{C\times 1\times 1}}$. For more details, please refer to the work of Woo et al. \cite{wooCBAMConvolutionalBlock2018}. The final channel attention ${M_l^c}$ can be written as:
\begin{equation}
	\begin{aligned}
		{M_l^c} &= \sigma ({F_{avg}} + {F_{\max }})\\
		&= \sigma (\mathbf{MLP}(\mathbf{Avg}({S_l^M})) + \mathbf{MLP}(\mathbf{Max}({S_l^M}))),
	\end{aligned}
	\label{eq:mc}
\end{equation}
where $\sigma$ stands for Sigmoid activation function.

The main idea of boundary-guided spatial attention is to predict segmentation probability maps based on high-level semantic features. By dilating and eroding of the segmentation probability maps, a ternary spatial confidence region can be obtained, which consists of high-confidence tumor regions, high-confidence background regions, and low-confidence lesion boundary regions. We use the ternary spatial confidence region to calculate boundary-guided spatial attention map and to guide semantic segmentation at the next level. Unlike the channel attention block, the input of the boundary-guided spatial attention block at the $l^{th}$ layer includes not only the output feature $S_{l}^{M}$ of the MGPM but also the output feature $D_{l+1}$ of the decoder at ${l+1}^{th}$ level since the decoded features can provide more accurate semantic information, which helps to calculate the ternary spatial confidence region more accurately. As shown in Fig. \ref{fig:bgm}, to match the size of the ${S_l^M}$ and ${D}_{l+1}$, we first upsample ${{D}_{l+1}}\in {{\mathbb{R}}^{64\times {{H}^{l+1}}\times {{W}^{l+1}}}}$ by a factor of 2 , then reduce its channels to 1 by a $1\times 1$ convolution and pass it through a sigmoid activation function to obtain a coarse segmentation probability map ${{P}_{l}}\in {{\mathbb{R}}^{1\times {{H}^{l}}\times {{W}^{l}}}}$:

\begin{equation}
	{P_l}=\sigma(\mathop{\mathbf{Conv}}\limits_ {1\times1}(\uparrow D_{l+1})) ,  \qquad l=1,2,3.
	\label{eq:p}
\end{equation}

The dilation operation expands the lesion area in the segmentation probability map $P_l$ to include as many border regions as possible. The erosion operation can be used to reduce the lesion area to obtain high-confidence tumor regions. Theorectically, the pixel values of high-confidence tumor regions in probability map are close to 1, while the pixel values of high-confidence background regions are close to 0, and due to the high uncertainty of regions near the tumor boundary, their pixel values are close to 0.5. Accordingly, the boundary-guided attention in this work is defined as:
\begin{equation}
	M_l^s=\left[ \mathop{\mathbf{Max}}\limits_{k_l^e}(P_l) + (-\mathop{\mathbf{Max}}\limits_{k_l^d}(-P_l))\right]/ 2,
	\label{eq:ms}
\end{equation}
where "$\mathbf{Max}$" indicates the max-pooling operation which is used to implement the dilation and erosion on the probability map, $k_{e}^{l}$ and  $k_{d}^{l}$ represent the size of the dilation and erosion kernels, respectively. 

After obtaining the output $S_l^B$ of the BGM module at the $l^{th}$ layer by combining  \eqref{eq:slb}-\eqref{eq:ms}, it is used as the input of the decoder at the $l^{th}$ layer. Finally, using the output of decoder $D_0$, the final segmentation prediction probability map is obtained after passing through the bottleneck layer and the Sigmoid function.

\subsection{Loss Function}
To further improve the boundary aware ability, in addition to the traditional segmentation loss $L^{seg}$, we also introduce a multi-level boundary-enhanced segmentation (BS) loss $L_l^{beseg}$. The total loss of the PBNet can be expressed as:

\begin{equation}	
	{L}_{total}=\sum\limits_{l=1}^{N}{\lambda}_{l}L_{l}^{beseg}(\Uparrow(P_l),G)+L^{seg}(P_0,G),
\end{equation}
where $P_0$ is the segmentation probability map obtained by the decoder feature $D_0$ through a Sigmoid function, $G$ is the true label of the segmentation, $P_l$ is defined as in \eqref{eq:p}, $\Uparrow(\cdot)$ means upsampling of the input to the same size as $G$, and ${{\lambda }_{l}}$ indicates the weight of the BS loss at different levels. In general, the deeper the layer, the more times it is downsampled, and the more blurred the boundary segmentation result is relative to it. Therefore, $\lambda_l$ should be set to a smaller value for deeper layers. The specific parameter settings are listed in the Table \ref{tab:kernelsize-setup}. 

\begin{table}[!t]
	\renewcommand\arraystretch{1.3}
	\centering
	\caption{Hyper-parameters setting for BGM and losses}
	\label{tab:kernelsize-setup}
	\resizebox{.8\linewidth}{!}{
		\begin{tabular}{cccccc}
			\hline\hline
			\multicolumn{2}{c}{$l$}               & 3 & 2 & 1 & 0 \\ \hline
			\multirow{2}{*}{BGM}  & $k_e$ & 3          & 5          & 7          & -          \\
			& $k_d$ & 5          & 9          & 13         & -          \\ \hline
			\multirow{2}{*}{Loss} & $k$  & 5          & 5          & 5          & -          \\
			& $ \lambda $  & 0.4        & 0.6        & 0.8        & 1          \\ \hline\hline
		\end{tabular}
	}
\end{table}

In this paper, the traditional segmentation loss $L^{seg}$ combines the weighted IoU loss and the Binary Cross Entropy (BCE) loss:
\begin{equation}	
	{L}^{seg}(a,b)=IoU(a,b)+BCE(a,b),
\end{equation}
where $a$ and $b$ represent any two inputs. 

The BS loss $L_l^{beseg}$ at the $l^{th}$ layer is composed of the segmentation loss and boundary-enhanced loss at that layer, namely:
\begin{equation}	
	{L}_l^{beseg}=\alpha_1 L^{seg}(\Uparrow(P_l),G) + \alpha_2 L^{boundary}(\Uparrow(P_l),G),
\end{equation}

To compute the boundary-enhanced loss $L^{boundary}$, we also adopt the concept of dilation and erosion. We implement the average pooling operation to enlarge the tumor region and then subtracted the original image to obtain the tumor boundary area. The boundary-enhanced loss is the mean squared error (MSE) between the actual boundary area and the predicted boundary area:
\begin{equation}	
	L^{boundary}(a,b)=MSE(\left|\mathop{\mathbf{Avg}}\limits_{k}(a)-a\right|, \left|\mathop{\mathbf{Avg}}\limits_{k}(b)-b\right|),
\end{equation}
where $k$ is the kernel size of average pooling and $\left|\cdot\right|$ stands for the absolute operation. We can control the weights of $L_{seg}$ and $L_{boundary}$ at each level by ${{\alpha }_{1}}$ and ${{\alpha }_{2}}$. In this work, we set ${{\alpha }_{1}}\text{:}{{\alpha }_{2}}=1:5$ to accelerate the learning progress on tumor boundaries.

\section{Experiments}
\label{sec:experiments}
\subsection{Data Description and preprocessing}
We train and test the proposed model on two datasets, namely the public dataset BUSI\cite{al-dhabyaniDatasetBreastUltrasound2020} and the in-house dataset. BUSI collects 780 ultrasound images with corresponding masks from 600 female patients, including 437 benign tumors, 210 malignant tumors, and 133 tumor-free cases. We resize all the images to $256 \times 256$. The in-house dataset contains a total of 995 ultrasound images, including 159 BIRADS-2 cases, 204 BIRADS-3 cases, 563 BIRADS-4 cases, and 69 BIRADS-5 cases. The experimental protocol was approved by the ethics committee of Guizhou Provincial People’s Hospital (No. (2023) 015). The breast tumor areas are all delineated by experienced radiologists, and we resize them into the same size of $256\times 384$.
During the experiment, all the images are normalized. Data augmentation strategies, such as flipping, rotation, and scaling implemented in the Albumentations\cite{info11020125}, has a 50\% chance to be executed in every epoch to avoid overfitting.

\subsection{Experimental setup}
To assess the effectiveness of our proposed method, we compared it with UNet\cite{olafUNetConvolutionalNetworks2015}, FPN\cite{tsung-yiFeaturePyramidNetworks2017}, DeepLabV3+\cite{chenEncoderDecoderAtrousSeparable2018}, FPNN\cite{tangFeaturePyramidNonlocal2021}, CaraNet\cite{louCaraNetContextAxial2022}, MSNet\cite{zhaoAutomaticPolypSegmentation2021}, and BDGNet\cite{qiuBDGNetBoundaryDistribution2022}. We utilize the implementation of UNet, FPN, and DeepLabV3+ in segmentation models Pytorch\cite{Iakubovskii:2019} and follow the official code provided for other models. Our model is implemented based on the PyTorch framework and trained on a single NVIDIA Tesla V100 GPU for 100 epochs with mini-batch size 16, and optimized with AdamW optimizer. The maximum learning rate is set to 0.001. Poly learning rate policy (with $power=0.9$) is used to adjust the learning rate. To fairly compare and verify the model’s generalization ability, we employ 5-fold cross-validation to test the their segmentation performance on each dataset.

\subsection{Evaluation metrics}
To conduct quantitative comparison, the Dice score (Dice), Jaccard coefficients (Jac),  Hausdorff Distance (HD)\cite{tahaEfficientAlgorithmCalculating2015}, Sensitivity (Sen), and specificity(Spe) are used to evaluate the segmentation performance. These metrics can be calculated as follows:
\begin{equation}Dice=\frac{2\times TP}{2\times TP+FP+FN},\end{equation}
\begin{equation}Jaccard=\frac{TP}{TP+FP+FN},\end{equation}
\begin{equation}Sen=\frac{TP}{TP+FN},\end{equation}
\begin{equation}Spe=\frac{TN}{TN+FP},\end{equation}
where TP (true positive) and TN (true negative) denote the number of pixels accurately predicted as foreground or background, respectively. FP (false positive) and FN (false negative) represent the number of pixels inaccurately predicted as foreground or background, respectively. In this study, the tumor region is taken as the foreground area (positive), while the remaining normal tissue region is taken as the background (negative).
HD measures the shape similarity between the ground truth and the segmentation result. The smaller the HD, the more similar the shapes of the ground truth and segmentation result. It is defined as:
\begin{equation}
	HD = \max (\mathop {\max }\limits_{y \in Y} \{ \mathop {\min }\limits_{g \in G} \{ \left\| {y - g} \right\|\} \} ,\mathop {\max }\limits_{g \in G} \{ \mathop {\min }\limits_{y \in Y} \{ \left\| {g - y} \right\|\} \} ),
	\label{eq:hd}
\end{equation}
where $y$ and $g$ represent any point in the segmentation result $Y$ and the segmentation label $G$, respectively, and$\left\| \cdot \right\|$ is the L2 norm distance metric function.

\section{RESULTS}
\label{sec:results}
\subsection{Comparison with SOTA methods}

\begin{table*}[!t]
	\renewcommand\arraystretch{1.3}
	\caption{Quantitative comparison results on BUSI dataset. The optimal and suboptimal results are highlighted with bold and underline, respectively.(Mean Value $\pm$ Standard Deviation)}
	\label{tab:busi}
	\centering
	\resizebox{0.8\textwidth}{!}{
	\begin{tabular}{ccccccc}
		\hline\hline
		\multicolumn{1}{l}{}          & \pmb{Models} & \pmb{Dice($\uparrow$)} & \pmb{Jac($\uparrow$)} & \pmb{Sen($\uparrow$)} & \pmb{Spe($\uparrow$)} & \pmb{HD($\downarrow$)}        \\ \hline
		\multirow{8}{*}{\pmb{BUSI}}  & UNet\cite{olafUNetConvolutionalNetworks2015}            & 79.3$\pm$2.76           & 70.94$\pm$3.02                & 83.91$\pm$2.64               & 97.53$\pm$0.29          & 5.08$\pm$0.25          \\
		& FPN\cite{tsung-yiFeaturePyramidNetworks2017}             & 77.87$\pm$3.48          & 68.99$\pm$4.36                & 83.99$\pm$2.23               & 97.25$\pm$0.43          & 5.27$\pm$0.39          \\
		& DeepLabV3+\cite{chenEncoderDecoderAtrousSeparable2018}      & 78.23$\pm$2.51          & 69.67$\pm$3.05                & 82.06$\pm$1.86               & \underline {97.75$\pm$0.53}    & 5.14$\pm$0.18          \\
		& FPNN\cite{tangFeaturePyramidNonlocal2021}            & 77.52$\pm$4.61          & 69.19$\pm$4.94                & 80.48$\pm$3.52               & 97.61$\pm$0.69          & 5.25$\pm$0.35          \\
		& MSNet\cite{zhaoAutomaticPolypSegmentation2021}  & {79.89$\pm$2.69} & {71.6$\pm$3.4}         & \underline {84.36$\pm$2.1} & 97.48$\pm$0.35          & 5.23$\pm$0.38          \\
		& CaraNet\cite{louCaraNetContextAxial2022}         & 79.7$\pm$2.48           & 71.49$\pm$3.0                 & 83.39$\pm$1.45               & 97.58$\pm$0.36          & 5.05$\pm$0.16          \\
		& BDGNet\cite{qiuBDGNetBoundaryDistribution2022}    & \underline {80.35$\pm$4.26}    &  \pmb{72.74$\pm$4.24} &  {83.87$\pm$4.25}         & 97.74$\pm$0.23          & \pmb{4.96$\pm$0.25} \\
		& \pmb{PBNet} & \pmb{81.1$\pm$2.21}  & \underline {{72.63$\pm$2.79}} & \pmb{85.56$\pm$1.23} & \pmb{97.75$\pm$0.32} & \underline {4.98$\pm$0.18}   \\ \hline
		\multirow{8}{*}{\pmb{BUSI*}} & UNet\cite{olafUNetConvolutionalNetworks2015}            & 81.16$\pm$2.72          & 72.75$\pm$3.06                & 86.18$\pm$3.1                & 97.55$\pm$0.46          & 4.95$\pm$0.19          \\
		& FPN\cite{tsung-yiFeaturePyramidNetworks2017}             & 81.06$\pm$3.33          & 72.67$\pm$3.75                & 85.78$\pm$2.73               & 97.55$\pm$0.21          & 4.9$\pm$0.2            \\
		& DeepLabV3+\cite{chenEncoderDecoderAtrousSeparable2018}      & 80.8$\pm$2.78           & 72.12$\pm$3.14                & 85.75$\pm$3.17               & 97.77$\pm$0.24          & 5.07$\pm$0.14          \\
		& FPN\cite{tangFeaturePyramidNonlocal2021}N            & 78.7$\pm$4.53           & 70.61$\pm$4.45                & 80.58$\pm$3.97               & 97.55$\pm$0.93          & 5.23$\pm$0.4           \\
		& MSNet\cite{zhaoAutomaticPolypSegmentation2021}           & \underline {82.38$\pm$2.72}    & \underline {74.31$\pm$3.32}          & 85.74$\pm$1.56               & \pmb{97.92$\pm$0.23} & \underline {4.82$\pm$0.23}    \\
		& CaraNet\cite{louCaraNetContextAxial2022}         & 80.94$\pm$3.85          & 72.33$\pm$4.67                & \pmb{87.1$\pm$1.88}       & 97.67$\pm$0.35          & 4.91$\pm$0.2           \\
		& BDGNet\cite{qiuBDGNetBoundaryDistribution2022}          & 81.33$\pm$4.96          & 73.72$\pm$4.73                & 84.98$\pm$5.45               & 97.42$\pm$0.5           & 4.95$\pm$0.35          \\
		& \pmb{PBNet} & \pmb{82.88$\pm$3.17} & \pmb{74.79$\pm$4.12}       & \underline {86.69$\pm$1.69}    & \underline {97.85$\pm$0.34}    & \pmb{4.75$\pm$0.2} \\ \hline\hline
	\end{tabular}
}
\end{table*}

\begin{table}[htbp]
	\centering
	\renewcommand\arraystretch{1.3}
	\caption{Quantitatively comparing PBNet with other sate-of-the-art methods on BUSI* dataset (\%)}
	\label{tab:busi*}
	\resizebox{\columnwidth}{!}{
		\begin{tabular}{cccccc}
			\hline\hline
			\pmb{Models} & \pmb{Params(M)} & \pmb{Dice($\uparrow$)} & \pmb{Jac($\uparrow$)} & \pmb{Sen($\uparrow$)} & \pmb{Spe($\uparrow$)}   \\
			\hline
			SegNet\cite{badrinarayananSegNetDeepConvolutional2017}          & 41.84           & 74.60          & 59.51          & 77.02          & 96.92          \\
			PSPNet\cite{zhaoPyramidSceneParsing2017}          & 49.07           & 75.48          & 60.63          & 77.74          & 97.01          \\
			Attention-UNet\cite{oktayAttentionUNetLearning2018}  & 35.58           & 71.35          & 55.50          & 74.21          & 96.37          \\
			STAN\cite{shareefStanSmallTumorAware2020}            & -               & 75.00          & 66.00          & 76.00          & -              \\
			ESTAN\cite{shareefESTANEnhancedSmall2020}           & -               & 78.00          & 70.00          & 80.00          & -              \\
			SAC\cite{huangShapeAdaptiveConvolutionalOperator2021}             & -               & 79.12          & 72.12          & 82.56          & -              \\
			MSSA-Net\cite{xuMssaNetMultiScaleSelfAttention2021}        & 71.53           & 80.65          & 71.90          & 81.06          & -              \\
			MSF-GAN\cite{huangMSFGANMultiScaleFuzzy2021}         & -               & 79.99          & 71.12          & 78.34          & -              \\
			MTL-COSA\cite{xuRegionalAttentiveMultiTaskLearning2023}        & 109.24          & 78.90          & 70.65          & 79.31          & \pmb{98.31} \\
			RMTL-Net\cite{xuRegionalAttentiveMultiTaskLearning2023}        & 93.51           & 80.04          & 71.93          & 82.54          & 98.00          \\
			BGRA-GSA\cite{huBoundaryguidedRegionawareNetwork2023}        & 101.34          & 81.43          & 68.75          & 84.14          & 97.63          \\
			\pmb{PBNet}  & \pmb{29.14}  & \pmb{82.88} & \pmb{74.79} & \pmb{86.69} & 97.85 \\    
			\hline\hline    
		\end{tabular}
	}
\end{table}

Table \ref{tab:busi} (BUSI) presents the quantitative evaluation metrics of BUS segmentation on the BUSI dataset for PBNet and other comparison models. PBNet achieves the highest average values for Dice, Sen and Spe, of 81.10\%, 85.56\% and 97,75\%, respectively. Compared to the sub-optimal model, Dice and Sen increase by 0.75\% and 1.2\%, respectively. To account for the potential influence of normal breast tissue ultrasound images on segmentation results, we remove all normal samples from the BUSI dataset to form the BUSI* dataset. The models are trained and validated on the BUSI* dataset, with results shown in the second panel of Table \ref{tab:busi} (BUSI*). To ensure comparability, tumor samples in each fold of the training and validation sets of BUSI* and BUSI remain consistent. Our proposed model achieves the highest average values for Dice and Sen on the BUSI* dataset, reaching 82.88\% and 86.69\%, respectively, and the smallest HD with an average value of 4.75 mm. 

To further illustrate the superiority of the proposed PBNet, we also compare our PBNet with several state-of-the-art image segmentation methods which are implemented on the same dataset (BUSI*), including SegNet\cite{badrinarayananSegNetDeepConvolutional2017}, PSPNet\cite{zhaoPyramidSceneParsing2017}, Attention-UNet\cite{oktayAttentionUNetLearning2018}, STAN\cite{shareefStanSmallTumorAware2020}, ESTAN\cite{shareefESTANEnhancedSmall2020}, SAC\cite{huangShapeAdaptiveConvolutionalOperator2021}, MSSA-Net\cite{xuMssaNetMultiScaleSelfAttention2021}, MSF-GAN\cite{huangMSFGANMultiScaleFuzzy2021}, MTL-COSA\cite{xuRegionalAttentiveMultiTaskLearning2023}, RMTL-Net\cite{xuRegionalAttentiveMultiTaskLearning2023}, and BGRA-GSA\cite{huBoundaryguidedRegionawareNetwork2023}. Note that, to avoid the bias caused by reimplementation, the quantitative results of all these comparing methods on the BUSI* dataset are directly derived from the corresponding papers. As shown in Table \ref{tab:busi*}, our approach significantly outperforms these methods in terms of Dice, Jac, and Sen, as well as the amount of model parameters. Compared to the sub-optimal BGRA-GSA, our PBNet has less than a third of its parameters, while achieving improvements of 1.45\%, 6.04\%, 2.55\%, and 0.22\% on the Dice, Jac, Sen, and Spe, respectively.

\begin{table*}[!t]
	\renewcommand\arraystretch{1.3}
	\caption{Quantitative comparison results on in-house dataset. The optimal and suboptimal results are highlighted with bold and underline, respectively.(Mean Value $\pm$ Standard Deviation)}
	\label{tab:inhouse}
	\centering
	\resizebox{0.8\textwidth}{!}{
	\begin{tabular}{cccccc}
		\hline\hline
		\pmb{Models} & \pmb{Dice($\uparrow$)} & \pmb{Jac($\uparrow$)} & \pmb{Sen($\uparrow$)} & \pmb{Spe($\uparrow$)} & \pmb{HD($\downarrow$)}        \\ \hline
		UNet\cite{olafUNetConvolutionalNetworks2015}       & 81.0$\pm$2.39       & 71.06$\pm$2.2        & 86.12$\pm$2.63          & 98.45$\pm$0.22       & 5.02$\pm$0.11      \\
		FPN\cite{tsung-yiFeaturePyramidNetworks2017}        & 81.27$\pm$2.38      & 71.33$\pm$2.29       & \pmb{87.84$\pm$2.08} & 98.21$\pm$0.32       & 5.05$\pm$0.12      \\
		DeepLabV3+\cite{chenEncoderDecoderAtrousSeparable2018} & 80.35$\pm$2.38      & 70.54$\pm$2.25       & 86.06$\pm$2.38          & 98.31$\pm$0.28       & 5.12$\pm$0.17      \\
		FPNN\cite{tangFeaturePyramidNonlocal2021}       & 77.96$\pm$2.02      & 67.59$\pm$2.0        & 82.36$\pm$0.95          & 98.47$\pm$0.32       & 5.3$\pm$0.16       \\
		MSNet\cite{zhaoAutomaticPolypSegmentation2021}      & 82.32$\pm$2.08      & 72.69$\pm$1.98       & 85.42$\pm$3.4           & \underline { 98.68$\pm$0.21} & 4.87$\pm$0.07      \\
		CaraNet\cite{louCaraNetContextAxial2022}    & \underline { 82.4$\pm$2.09} & \underline { 72.77$\pm$1.95} & 85.64$\pm$1.76          & 98.67$\pm$0.33       & 4.9$\pm$0.11       \\
		BDGNet\cite{qiuBDGNetBoundaryDistribution2022}     & 82.25$\pm$2.28      & 72.71$\pm$2.18       & 85.64$\pm$2.33          & 98.65$\pm$0.23       & \underline { 4.83$\pm$0.1} \\
		\pmb{PBNet}  & \pmb{83.11$\pm$1.96} & \pmb{73.8$\pm$1.74} & \underline { 86.32$\pm$1.56} & \pmb{98.77$\pm$0.27} & \pmb{4.75$\pm$0.13} \\ \hline\hline
	\end{tabular}}
\end{table*}

\begin{figure*}[!t]
	\centering
	\includegraphics[width=0.8\linewidth]{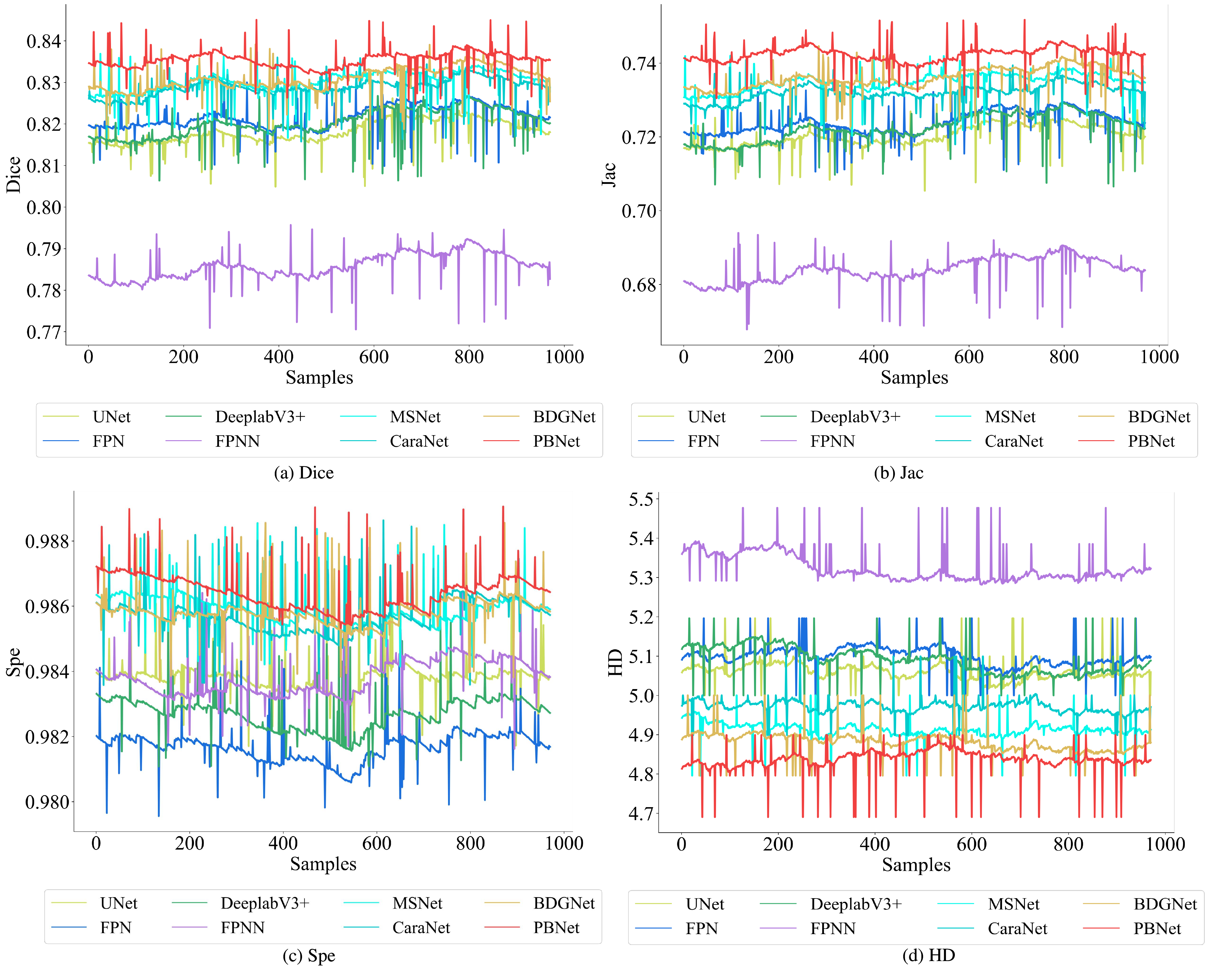}
	\caption{Quantitative metric curves of comparison models on the in-house dataset. The curves in (a), (b), (c), and (d) with different colors indicate Dice, Jac, Spe, and HD values of different models, respectively.}
	\label{fig:curve}
\end{figure*}

To test the generability of PBNet, we also perform comparative experiments on in-house dataset. The results are presented in Table \ref{tab:inhouse}. PBNet achieves significant improvements in almost all evaluation metrics (except Sen) compared to other models. Relative to the sub-optimal model, Dice, Jac and Spe improve by approximately 0.7\%, 1.1\%, and 0.1\%, respectively, while HD decreases by 1.7\%. To evaluate the segmentation performance of each model more comprehensively on different samples, we plot the Dice, Jac, Sen, and HD curves (within 95\% confidence interval) for all samples of in-house dataset. As shown in Fig. \ref{fig:curve}, FPNN has relatively poor segmentation performance on almost all samples and exhibits low stability for different samples (the metric fluctuates significantly with sample changes). Although the stability of other models is comparable to that of the proposed PBNet, PBNet has the best overall performance on all metrics for all samples compared to them.

\begin{figure*}[!t]
	\centering
	\includegraphics[width=0.8\linewidth]{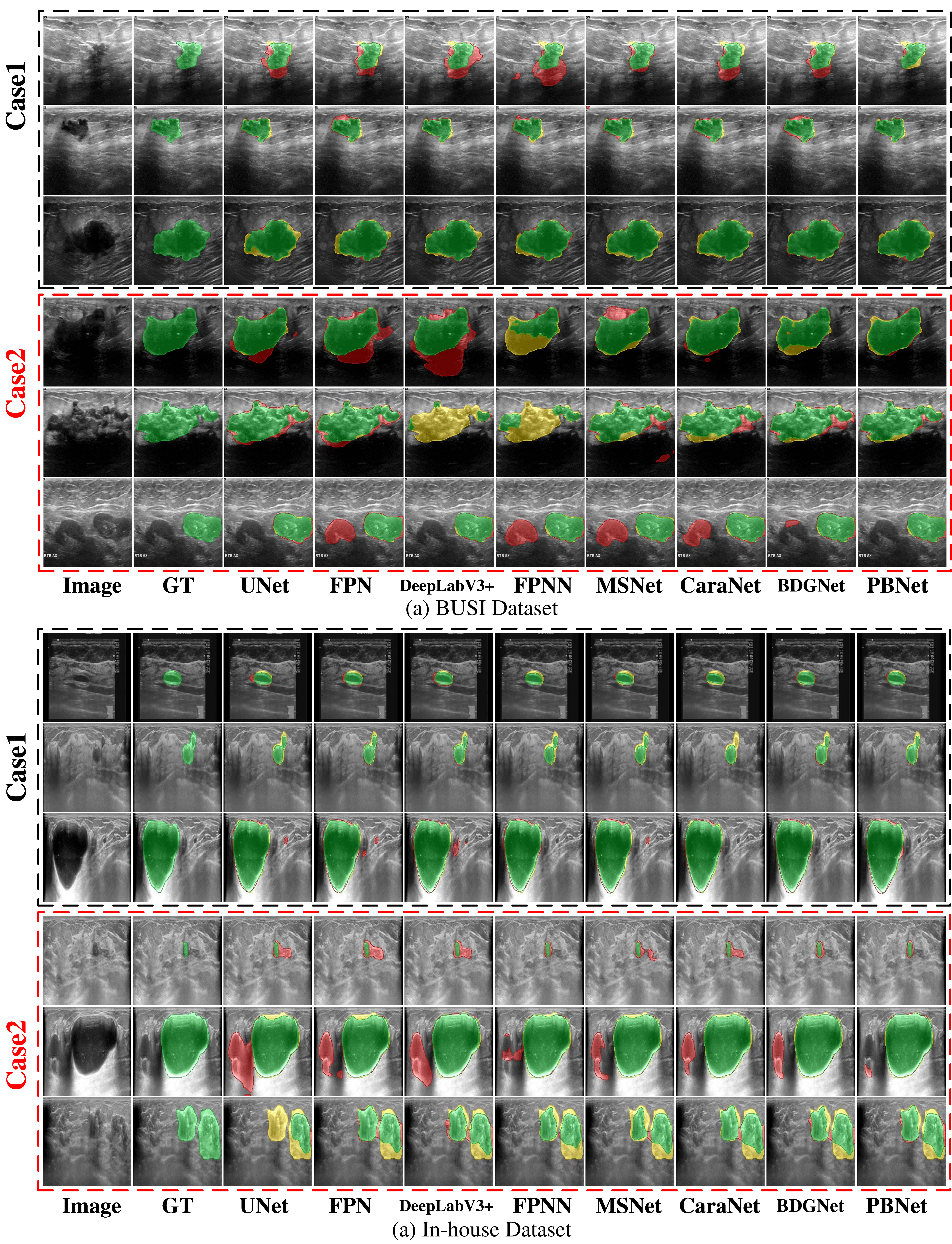}
	\caption{Qualitative assessment of segmentation predicted by different models on BUSI and the in-house dataset. The green area represents the true region, yellow represents the under-segmented region, and the red area represents the over-segmented region.}
	\label{fig:seg}
\end{figure*}

To qualitatively compare the segmentation performance of different models on different samples, Fig. \ref{fig:seg}(a) and (b) visualize the segmentation results for the tumors with different properties in both BUSI and in-house datasets, respectively. For better visualization, all images have been resized to the same size. Among them, the green area represents the true region, the yellow area represents the under-segmented region (i.e., the tumor is missed), and the red area represents the over-segmented region (the normal tissue is mistaken for the tumor). It can be observed that on the BUSI dataset, for tumor regions with regular shapes and high contrast (Case 1), all models can recognize the tumor region well. However, compared to other models, there are fewer over-segmentation and under-segmentation phenomena in the segmentation results of the proposed model, and the segmentation performance is better at the edge of the tumor. For images with irregular shapes and low contrast (Case 2), UNet, FPN, DeepLabV3+ and FPNN cannot segment well with many tumor regions being missed, whereas MSNet, CaraNet, BDGNet and our model can handle this situation well. Additionally, when a large low-echogenic area is connected to the tumor region, most models struggle to distinguish them, resulting in poor segmentation results. Although the proposed model can distinguish them, the segmentation performance is slightly decreased compared to the high-contrast tumor regions. On the in-house dataset, we also observed the same pattern, that means for tumors with irregular shapes and slightly lower contrast, the superiority of the proposed model is more obvious.

\subsection{Ablation}

\subsubsection{Module ablation}
\begin{table*}[t]
	\renewcommand\arraystretch{1.3}
	\caption{Quantitative metrics for ablation experiments.(Mean Value $\pm$ Standard Deviation)}
	\label{tab:ablation results}
	\centering
	\resizebox{0.8\textwidth}{!}{
	\begin{tabular}{cccccccccc}
		\hline\hline
		\pmb{MGPM} & \pmb{BGM} & \pmb{BS} & \pmb{MACs(G)} & \pmb{Params(M)} & \pmb{Dice($\uparrow$)} & \pmb{Jac($\uparrow$)} & \pmb{Sen($\uparrow$)} & \pmb{Spe($\uparrow$)} & \pmb{HD($\downarrow$)} \\ \hline
		&   &   & 5.57 & 28.77 & 79.63$\pm$3.28 & 71.32$\pm$3.63 & 84.93$\pm$3.06 & 97.38$\pm$0.27 & 5.18$\pm$0.27 \\
		\checkmark &   &   & 5.96 & 28.91 & 80.46$\pm$2.98 & 72.06$\pm$3.32 & 85.21$\pm$2.55 & 97.4$\pm$0.42  & 5.01$\pm$0.18 \\
		& \checkmark &   & 7.17 & 29.0  & 80.25$\pm$2.24 & 71.82$\pm$2.97 & 84.3$\pm$2.61  & 97.57$\pm$0.46 & 5.07$\pm$0.14 \\
		& \checkmark & \checkmark & 7.17 & 29.0  & 80.68$\pm$3.61 & 72.21$\pm$4.06 & 85.65$\pm$3.36 & 97.43$\pm$0.22 & 5.14$\pm$0.25 \\
		\checkmark & \checkmark & \checkmark & 7.56  & 29.14 & 81.1$\pm$2.21  & 72.63$\pm$2.79 & 85.56$\pm$1.23 & 97.75$\pm$0.32 & 4.98$\pm$0.18 \\ \hline\hline
	\end{tabular}}
\end{table*}

To evaluate the effectiveness of the different modules proposed in this paper, we design different ablation experiments on the BUSI dataset. The results are shown in Table \ref{tab:ablation results}. After the introduction of MGPM, all metrics are comprehensively improved with only a 0.14 M increase in the number of parameters. Among them, Dice, Jac, sen, and Spe improve by 0.83\%, 1.12\%, 0.28\%, and 0.02\% compared to baseline, and HD decreases by 0.17 mm. After introducing BGM with a 0.23M increase in the number of parameters, Dice, Jac, Spe, and HD improve by 0.62\%, 0.5\%, 0.19\%, and 2.1\%, respectively, but Sen decreases slightly. After introducing BS loss based on BGM, we find that the model performance is further improved. Compared to the Baseline, Dice, Jac, and Sen improve by about 1.1\%, 0.9\%, and 0.7\%, respectively. Finally, after introducing all the modules, the segmentation performance of our PBNet reaches its best.

\begin{figure}[!t]
	\centering
	\includegraphics[width=0.8\columnwidth]{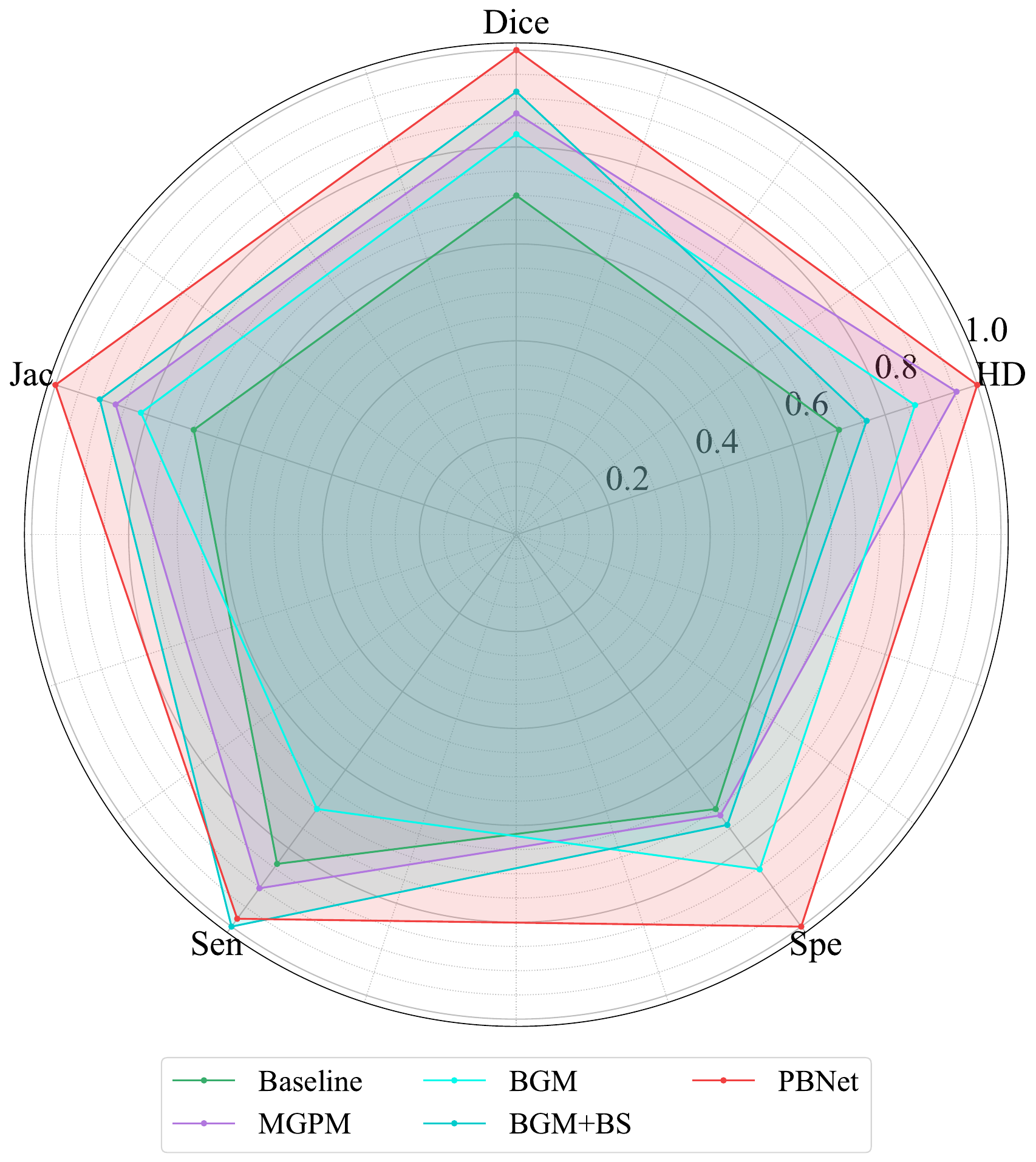}
	\caption{Radar plot of Dice, Sen, Jac, Sen, and HD of different ablation models on BUSI dataset.}
	\label{fig:radar}
\end{figure}

To intuitively demonstrate the effectiveness of each module, we normalize the segmentation metrics of the ablation experiments and display the radar plot in Fig. \ref{fig:radar}. It can be observed that after the addition of the MGPM alone, all metrics have improved compared to the Baseline. After adding the BGM, although the improvement of Dice, Sen, and Jac is not as pronounced as adding MGPM, after incorporating the BS loss, various evaluation metrics can be further improved and exceed the effect of adding the MGPM. After introducing all the modules simultaneously, the overall performance of the model is the best.

\begin{figure*}[!]
	\centering
	\includegraphics[width=0.8\linewidth]{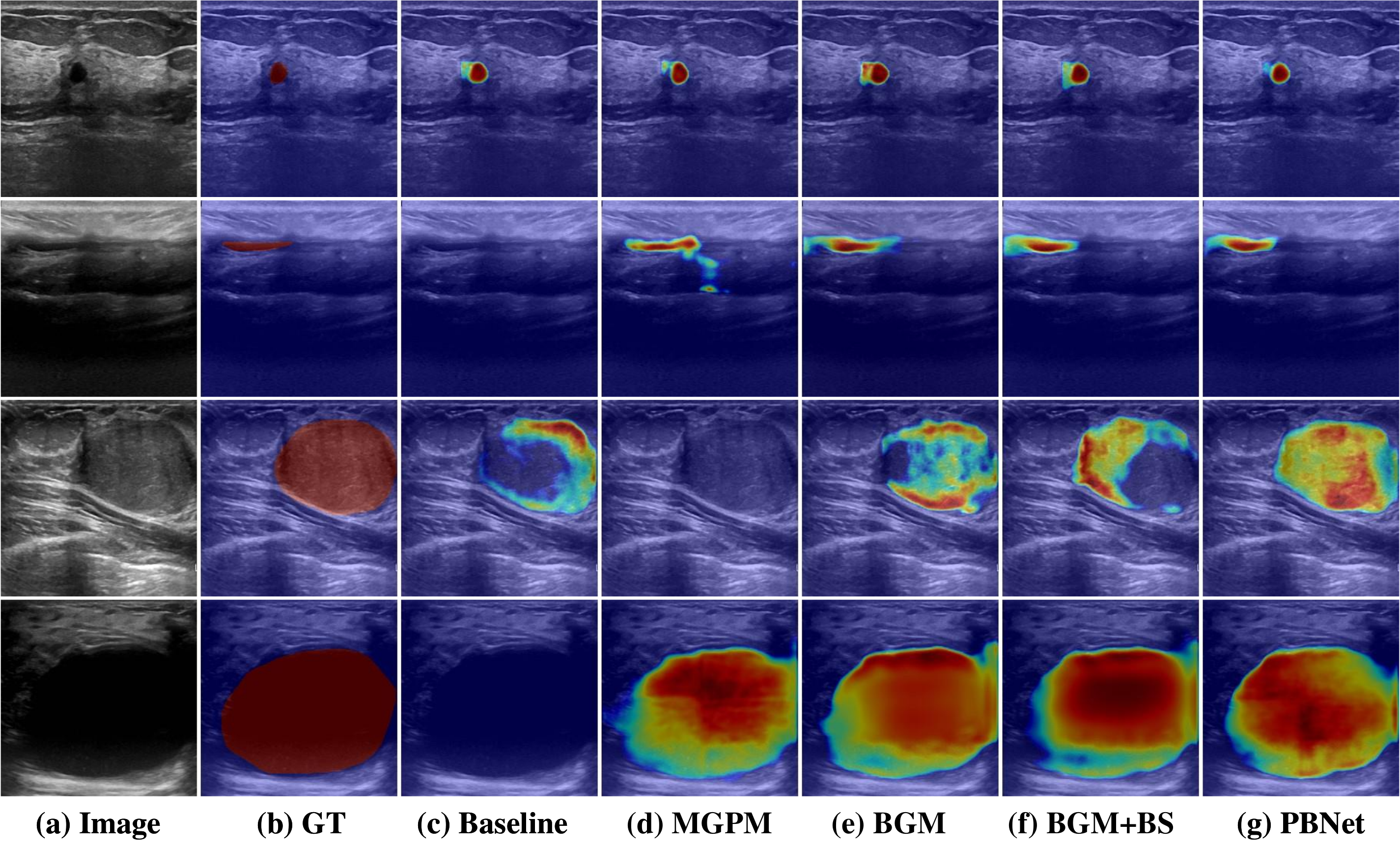}
	\caption{Comparison in attention maps obtained with models of different modules. The darker the red color of a region, the more attention the model pays to that region.}
	\label{fig:attention}
\end{figure*}

To further demonstrate the effectiveness of each module, we visualize the attention maps at the first layer of the individual ablation models. As shown in Fig. \ref{fig:attention}, with the help of powerful context information capture capability of MGPM, when the tumor connects with the low-echogenic regions, the model can better focus on the substantive tumor area and improve its ability to distinguish low-echogenic regions. 
Simultaneously, by using the ternary spatial confidence region in BGM to correct attention map,  the tumor boundary becomes more accurate. Compared to the Baseline, the attention map achieved by the proposed model is more uniform and covers almost all tumor areas.

\begin{figure}[!t]
	\centering
	\includegraphics[width=1\columnwidth]{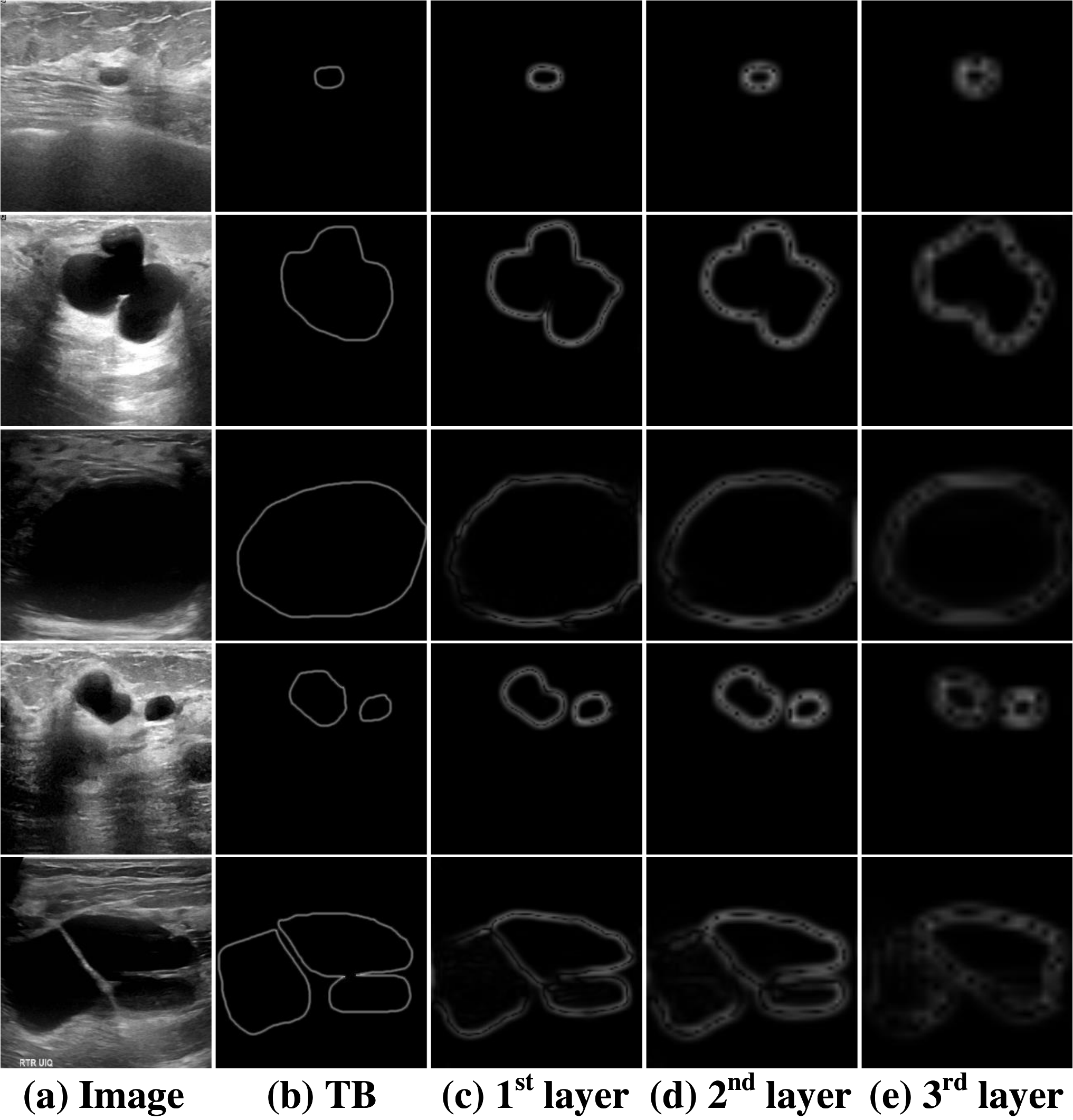}
	\caption{Boundaries obtained with BGM at different layers. TB represents the true boundary map obtained from the real segmentation label, the smaller the number of layers, the closer the prediction boundaries are to the final output and therefore more refined.}
	\label{fig:boundaries}
\end{figure}

Additionally, we visualize the boundary detection results of BGM at each layer in Fig. \ref{fig:boundaries}, where TB represents the true boundary obtained from the real segmentation label through the dilation and erosion operations. The ${l^{st}}$ layer is the output of BGM closest to the final output. After multiple layers of BGM operations, due to the combination of accurate tumor information from the decoder and detailed spatial information from the encoder, the boundary detection results become increasingly detailed and closer to TB. Even in cases of small tumor areas, unclear tumor edges, and multiple target areas, our BGM can detect tumor boundaries effectively. These results fully demonstrate the effectiveness of our BGM.

\subsubsection{Comparing MGPM with the similar modules}

\begin{table*}[!t]
	\renewcommand\arraystretch{1.3}
	\caption{Quantitative comparisons among ASPP, MGPM, CBAM, and BGM. MACs and Params indicate the increase in value after each module is added.(Mean Value $\pm$ Standard Deviation))}
	\label{tab:aspp_cbam}
	\centering
	\resizebox{0.8\textwidth}{!}{
	\begin{tabular}{cccccccc}
		\hline\hline
		\pmb{Components} & \pmb{MACs(G)} & \pmb{Params(M)} & \pmb{Dice($\uparrow$)} & \pmb{Jac($\uparrow$)} & \pmb{Sen($\uparrow$)} & \pmb{Spe($\uparrow$)} & \pmb{HD($\downarrow$)}\\ \hline
		ASPP\cite{chenDeepLabSemanticImage2018}   & 2.93  & 0.42 & 79.11$\pm$3.83 & 70.75$\pm$4.42 & 83.7$\pm$3.48  & 97.36$\pm$0.42 & 5.29$\pm$0.38 \\
		MGPM   & 0.39 & 0.14 & 80.46$\pm$2.98 & 72.06$\pm$3.32 & 85.21$\pm$2.55 & 97.4$\pm$0.42  & 5.01$\pm$0.18 \\ \hline
		CBAM\cite{wooCBAMConvolutionalBlock2018}   & 1.59 & 0.23  & 79.97$\pm$2.89 & 71.61$\pm$3.46 & 84.51$\pm$3.1  & 97.67$\pm$0.28 & 5.07$\pm$0.22 \\
		BGM+BS & 1.60 & 0.23  & 80.68$\pm$3.61 & 72.21$\pm$4.06 & 85.65$\pm$3.36 & 97.43$\pm$0.22 & 5.14$\pm$0.25 \\ \hline\hline
	\end{tabular}}
\end{table*}

The function of MGPM is like that of ASPP\cite{chenDeepLabSemanticImage2018} in terms of the improvement of the network's ability to capture contextual information. We compare the performance of the model on the BUSI dataset after introducing ASPP and MGPM. Table \ref{tab:aspp_cbam} shows that the introduction of ASPP increases the number of model parameters by 0.42M and the MACs by 2.93G. However, compared to the baseline, the introduction of ASPP caused a severe decrease in model performance. Since BGM is inspired by CBAM\cite{wooCBAMConvolutionalBlock2018}, we also compare how the model performs on the BUSI dataset after introducing CBAM and BGM. Table \ref{tab:aspp_cbam} shows that although the two modules have almost the same performance cost, the performance of BGM far exceeds that of CBAM. For Dice, Jac, and Sen, introducing BGM improves model performance by 0.71\%, 0.60\%, and 1.14\%, respectively, compared to introducing CBAM.

\section{Discussion}
\label{sec:discussion}
Accurate, automatic, and real-time breast tumor segmentation from BUS images is very significant for subsequent clinical analysis, treatment planning, and prognostic evaluation. However, due to the influence of low-echogenic areas and the high variability in the breast tumor shape and size, segmenting accurately the heterogenous breast lesions from BUS images is still a challenge. A large number of works have already demonstrated the effectiveness of UNet and its variants in the task of BUS image segmentation. However, considering the large semantic gap between the features of encoder and decoder in UNet, simply merging them by skip connections makes it difficult to effectively transfer semantic information. Therefore, such models do not perform well for the tumors with low image contrast. In addition, continuous convolution and pooling operations in UNet can loss the image details which cannot be recovered by simply upsampling, resulting in the worse segmentation results on lesion boundaries. To address these issues, we propose a PBNet in this work by introducing MGPM and BGM modules in the UNet skip connections to improve the segmentation results of tumors with low-contrast and irregular boundaries. By comparing with several state-of-the-art methods on two datasets, we validate the superiority of the proposed model.

In MGPM, it integrates the local and distant information in both intra- and inter feature maps with different semantic levels, which allows capturing the long-range dependency and thus bridging the semantic gap between encoder and decoder feature maps. The consistency in semantics of encoder and decoder can enhance the feature expression ability for tumors, enabling the model to tell the low-echogenic normal areas and tumor regions apart, or tell the normal tissues and low-contrast tumors apart. The ablation results present in Table \ref{tab:ablation results} and Fig. \ref{fig:attention} also validate this point. In Table \ref{tab:ablation results}, with the help of MGPM, both sensitivity and specificity are increased, which illustrates that the false negative and false positive are decreased simultaneously, demonstrating the ability of MGPM in distinguishing tumors and normal tissues. From the third and fourth rows of Fig. \ref{fig:attention} (a,c,d), we also observe that with MGPM, the model can focus on the tumor regions no matter how much the tumor area resembles the normal regions. The role of MGPM is similar to that of ASPP, however, we find that ASPP perform worse (Table \ref{tab:ablation results}) since it has a larger number of parameters rendering it more susceptible to overfitting for medical image dataset with smaller size. 

After MGPM, the BGM is also introduced into the skip connections. BGM and BS loss can guide the network to pay more attention on the boundary area, therefore improving the boundary segmentation performance. From the visualization results in Fig. \ref{fig:seg}, we can see that our proposed PBNet can segment tumor edges more accurately than other networks. This can also be reflected by the ablation results in Table \ref{tab:ablation results}, introducing BGM and BS loss can refine the segmented tumor boundaries, with the HD decreasing from 5.01 to 4.98 mm, Dice, Jac, Sen and Spe all being increased. This illustrates that the boundary-guided attention used in BGM can guide the next decoder layer to focus on the tumor edges. Comparing against other boundary-guided attentions, such as CBAM in Table \ref{tab:aspp_cbam}, we find that BGM is much better since it uses deep semantic features to calculate boundary-guided spatial attention, which may provide effective guidance for the model to be aware of the boundaries.  

With the help of both MGPM and BGM, the proposed PBNet outperforms other state-of-the-art models on both datasets. From Table \ref{tab:busi} and Table \ref{tab:inhouse}, we notice that the standard deviation of almost all the evaluation metrics calculated by PBNet is the smallest, indicating its stability and robustness. Compared the results of the models trained with BUSI (all samples) and BUSI* (only tumor samples) datasets in Table \ref{tab:busi}, we also observe that FPN, DeepLabV3+, and MSNet show significant increases in Dice scores, Jac coefficients, and Sen when trained with BUSI* dataset. However, CaraNet, BDGNet, and PBNet show smaller improvements, indicating they are less affected by normal samples and can more accurately distinguish tumor areas. Furthermore, we can see from Table \ref{tab:busi*} that even with a significantly fewer amount of parameters than other models, PBNet still achieves optimal segmentation performance, demonstrating feasibility of lightweight model in accurate segmentation of breast tumors.

Even though the proposed PBNet can distinguish the normal tissues that resemble the tumors, and to detect the tumor boundaries more accurately, there are still some limitations. For instance, the wrong segmentation still readily occurs when the tumor region is too small, or there is no obvious boundary between tumors and normal tissues. Additionally, when multiple tumor regions are present in one image, the PBNet can only identify some tumor regions and is unable to perform complete and accurate segmentation for all the tumors. How to deal with the segmentations for multiple tumors and small tumors is one of our future research interests. Moreover, the proposed BS loss was implemented at multi-levels, it is difficult to adjust their weights at different levels, adaptively and automatically setting the multilevel loss weights in the future work may bring more improvements.

\section{Conclusion}
\label{sec:conclusion}
In this work, we propose a novel breast lesion segmentation network, PBNet, which allows us to segment the non-enhanced lesions with more accurate boundaries with the help of multiple global perception module (MGPM) and boundary guided module (BGM), as well as the multi-level boundary-enhanced segmentation (BS) loss. MGPM adaptively fuse the long-range spatial information at both low and high-level feature maps, which can reduce the semantic gap between the features of encoder and decoder and promote the recognition ability for non-enhanced tumors. BGM uses the boundary-aware spatial attention derived from the high-level features to guide the model to focus on the boundary regions when extracting low level features. Moreover, the boundary-aware loss can also promote the boundary segmentation accuracy. The PBNet achieves the highest Dice score on both public and in-house datasets, which outperforms the existing state-of-the-art ultrasound segmentation models.

\section*{References}
\bibliographystyle{IEEEtran}
\bibliography{ref.bib}

\end{document}